\documentclass[reqno,11pt]{amsart}
\usepackage{amsmath,amssymb,tabularx,setspace,color,enumerate}

\newtheorem{corollary}{Corollary}%[section]
\newtheorem{theorem}{Theorem}
\usepackage{multirow}
\usepackage{array}
\usepackage{tabularx}
\usepackage{geometry}
\usepackage{amsmath}
\usepackage{setspace}
\usepackage{amsthm}
\usepackage{graphicx}
\usepackage{enumerate}
\makeatletter
\renewcommand\section{\@startsection {section}{1}{\z@}%
                                   {-3.5ex \@plus -1ex \@minus -.2ex}%
                                   {2.3ex \@plus.2ex}%
                                   {\normalfont\large\bfseries}}
\makeatother
\begin{document}
\doublespace
\title[]{A new goodness of fit test for   gamma distribution with censored observations}
\author[]%
{V\lowercase{aisakh} K. M.$^{\lowercase{a}}$, S\lowercase{reedevi} E.  P.$^{\lowercase{b}}$  \lowercase{and }  S\lowercase{udheesh} K. K\lowercase{attumannil$^{\lowercase{c},\dag}$}\\
$^{\lowercase{a}}$S\lowercase{t.} T\lowercase{homas} C\lowercase{ollege}, T\lowercase{hrissur,} K\lowercase{erala}, I\lowercase{ndia},\\
$^{\lowercase{b}}$SNGS C\lowercase{ollege,} P\lowercase{attambi,} K\lowercase{erala}, I\lowercase{ndia},\\
$^{\lowercase{c}}$I\lowercase{ndian} S\lowercase{tatistical} I\lowercase{nstitute},  C\lowercase{hennai}, I\lowercase{ndia}.
}
\thanks{ {$^{\dag}$}{Corresponding author E-mail: \tt skkattu@isichennai.res.in}.}
\maketitle
%(A\lowercase{utonomous})
 \begin{abstract}
 In the present paper, we develop a new goodness fit test for gamma distribution using the fixed point characterization. U-Statistic theory is employed to derive the test statistic. We discuss how the right censored observations are incorporated in the test developed here. The asymptotic properties of the test statistic in both censored and uncensored cases are studied in detail. Extensive Monte Carlo simulation studies are carried out to validate the performance of the proposed tests. We also illustrate the test procedure using several real data sets.
\end{abstract}

\section{Introduction}

Lifetime data analysis involve the modeling of time to event data. Several parametric distributions are used to model lifetime data. Exponential, gamma and Weibull distributions are some commonly used lifetime models. In this context, it is important to check the validity of an assumed parametric model. The goodness of fit tests are employed to validate the assumption that the lifetime data follows a particular distribution. A detailed study of goodness of fit tests for lifetime data is given in Lawless (2011). Gamma distribution has a great significance in lifetime distributions, due to its ability in modelling different ageing patterns. Gamma distribution  generalises exponential, $\chi^2$ and Erlang distributions. For applications of gamma distribution in lifetime data analysis one can refer Barlow and Proschan (1996) and Deshpande and Purohit (2015) among many others. Widespread applications of gamma distribution include modeling of rainfall data in Africa (Husak et al., 2007) and vinyl chloride data from an environmental study (Bhaumik et al., 2009) along with others.

A number of goodness of fit tests are available for gamma distribution in literature. % Assume that we have $n$ independent and identically distributed (i.i.d.) observations $X_1,...,X_n$ on a positive valued random variable $X$. Our interest is to test whether the distribution of $X$ belongs to $\Gamma(k,\lambda)$, where $k>0$ is the scale parameter and $\lambda>0$ is the shape parameter of gamma distribution.
    Kallioras et al. (2006) proposed a method using the empirical moment generating function to develop goodness of fit for gamma distribution. Henze et al. (2012) developed a goodness of fit test for gamma distribution based on empirical Laplace transform.   Villaseñor and González-Estrada (2015) suggested a variance ratio test for testing gamma distribution.   Baringhaus et al. (2017) proposed tests based on some independent properties of the gamma distribution. Recently, Betsch and Ebner (2019) developed a new characterisation for gamma distribution and an associated goodness of fit test. Note that all these tests  are developed for complete data.

Stein (1972) introduced a natural identity for a random variable whose distribution belongs to an exponential family. Stein’s identity and its role in inference procedures have been discussed widely in the literature. For a detailed discussion on Stein’s type identity for a general class of probability distributions and related characterizations, one can refer to Sudheesh (2009) and Sudheesh and Dewan (2016) and the references therein.   % Ross (2011) discussed approximations of the Normal, Poisson, Exponential and Geometric distributions using Stein’s method.
%As a special case,  let $X$ be a continuous random variable with finite mean $\mu$ and variance $\sigma^2$. Let $c(x)$ be a continuous function having first derivative. Then $X$ has normal distribution with mean $\mu$ and variance $\sigma^2$ if and only if
%\begin{equation*}\label{stein}
%E(c(X)(X-\mu))=\sigma^2E(c'(X)),
%\end{equation*}
%This has come to be known in literature as Stein’s identity or Stein’s lemma.
Using Stein’s type identity, Betsch and Ebner (2019) developed a fixed point characterization for gamma distribution. Making use of this characterization, we develop a U-statistic based goodness fit test for gamma distribution  for complete data. We also develop a new goodness fit test for gamma distribution with censored observations.

The rest of the paper is organized as follows. In Section 2, we develop a new non-parametric test for gamma distribution for complete data. In Section 3, we discuss how to incorporate right-censored observations in the testing procedure. We obtain  the asymptotic distribution of these test statistics separately. The result of Monte Carlo simulation studies are reported in Section 4 to evaluate the finite sample performance of the proposed tests. The procedures are illustrated with real data sets in Section 5. Finally, Section 6 concludes the study with a discussion on future works.
\section{Test statistics: Uncensored case}
In this section, we develop a new  goodness of fit test for gamma distribution with complete data. Let $X$ be non-negative random variable having distribution function $F$. Then $X$ has gamma distribution with parameter $k$ and $\lambda$ (denoted as $\Gamma(k,\lambda)$), if its probability density function is given by
\begin{equation*}\label{gammad}
  f(x)=\frac{\lambda^{-k}}{\Gamma(k)}x^{k-1}e^{-x/\lambda}, ~~~~~~x>0.
\end{equation*}
 We use the fixed point characterization based on Steins's type identity for gamma distribution to develop the test.
\begin{theorem}[Betsch and Ebner, 2019]
The random variable $X$ has gamma distribution with parameter $k$ and $\lambda$ if and only if
$$F(t)=E\left[\left(\frac{1-k}{X}+\frac{1}{\lambda}\right) \min(X,t)\right].$$
\end{theorem}
Based on a random sample $X_{1}, ...,X_{n}$  from $F$, we are interested in  testing the null hypothesis
$$H_{0}:F \in {\Gamma(k,\lambda)}$$
against$$ H_1: F \notin {\Gamma(k,\lambda)}.$$
 For testing the above hypothesis, we define a departure measure which discriminate between null and alternative hypothesis. Consider  $ \Delta(F)$ given by
\begin{eqnarray}\label{deltam}
\Delta(F)&=&\int_{0}^{\infty}\left(E\left[\left(\frac{1-k}{X}+\frac{1}{\lambda}\right)\min(X,t)\right]-F(t)\right)dF(t).
\end{eqnarray}In view of Theorem 1, $\Delta(F)$ is zero under  $H_0$ and non zero under $H_1$. Hence $\Delta(F)$  can be considered as a measure of departure  from the null hypothesis $H_0$ towards the alternative  hypothesis $H_1$.
As we propose the  test using the theory of U-statistics, first we simplify $\Delta(F)$ in terms of expectation of the function of random variables.
Consider
\begin{eqnarray}\label{delta}
\Delta(F)&=&\int_{0}^{\infty}E\left[\left(\frac{1-k}{X}+\frac{1}{\lambda}\right)\min(X,t)\right]dF(t)-\int_{0}^{\infty}F(t)dF(t).\nonumber\\
&=&\int_{0}^{\infty}\int_{0}^{\infty}\left(\frac{1-k}{x}\min(x,t)+\frac{1}{\lambda}\min(x,t)\right)dF(x)dF(t)- 1/2\nonumber\\&=&\Delta_{1}+\frac{1}{\lambda}E(\min(X_{1},X_{2}))-\frac{1}{2},
\end{eqnarray}where $\Delta_{1}=\int_{0}^{\infty}\int_{0}^{\infty}\frac{1-k}{x}\min(x,t)dF(x)dF(t).$ Now
\begin{eqnarray}\label{delta1}
\Delta_{1}&=&\int_{0}^{\infty}\int_{0}^{\infty}\frac{1-k}{x}\min(x,t)dF(x)dF(t).\nonumber\\
&=&(1-k)\int_{0}^{\infty}\int_{0}^{\infty}\frac{1}{x}\min(x,t)dF(x)dF(t).\nonumber\\
&=&(1-k)\int_{0}^{\infty}\left(\int_{0}^{\infty}I(x<t)+\frac{t}{x}I(t<x)\right) dF(x)dF(t).\nonumber\\
&=&(1-k)P(X_{1}<X_{2})+(1-k)\int_{0}^{\infty}\int_{t}^{\infty}\frac{t}{x}dF(x)dF(t).\nonumber\\
&=&(1-k)\frac{1}{2}+(1-k)E\left[\frac{X_{1}}{X_{2}}I(X_{1}<X_{2})\right],
\end{eqnarray}where $I(A)$ denote the indicator function of a set $A$. Substitute equation (\ref{delta1})  in equation (\ref{delta}) we obtain
\begin{eqnarray}\label{deltamnew}
\Delta(F)&=&\frac{1-k}{2}-\frac{1}{2}+(1-k)E\left[\frac{X_{1}}{X_{2}}I(X_{1}<X_{2})\right]+
\frac{1}{\lambda}E(\min(X_{1},X_{2}))\nonumber\\
&=&\frac{1}{\lambda}E(\min(X_{1},X_{2}))+(1-k)E\left(\frac{X_{1}}{X_{2}}I(X_{1}<X_{2})\right)-\frac{k}{2}.
\end{eqnarray}
We find test statistic using  U-statistics theory. Define $h_1(X_1,X_2)=\min(X_{1},X_{2})$. Then a U-statistic defined by
$$U_1=\frac{2}{n(n-1)}\sum_{i=1}^{n}\sum_{j=1,j<i}^{n}h_1(X_i,X_j),$$ is an unbiased estimator of  $E(\min(X_{1},X_{2}))$. Defined a symmetric kernel
$$h_2(X_1,X_2) =\frac{1}{2} \left(\frac{X_{1}}{X_{2}}I(X_{1}<X_{2})+\frac{X_{2}}{X_{1}}I(X_{2}<X_{1})\right) .$$ Then a U-statistic defined by
$$U_2=\frac{2}{n(n-1)}\sum_{i=1}^{n}\sum_{j=1,j<i}^{n}h_2(X_i,X_j),$$ is an unbiased estimator of $E\left(\frac{X_{1}}{X_{2}}I(X_{1}<X_{2})\right)$.
 Let $\widehat{\lambda}$ and $\widehat k$ be the consistent estimators of ${\lambda}$ and $k$, respectively.
Hence the test statistic is given by
\begin{equation}
 \widehat{\Delta} =\frac{U_{1}}{\widehat{\lambda}}+(1-\widehat{k})U_{2}-\frac{\widehat k}{2}.
\end{equation}
We reject the null hypothesis $H_0$ against the alternative  $H_1$ for large value of $\widehat{\Delta}$.

Next we study the asymptotic properties of the test statistic.  Since $U_1$ and $U_2$ are U-statistics they are consistent estimators of $E(\min(X_{1},X_{2}))$ and $E\left(\frac{X_{1}}{X_{2}}I(X_{1}<X_{2})\right)$, respectively (Lehmann, 1951). Hence  the following result is straight forward.
\begin{theorem}Under $H_1$, as $n\rightarrow \infty$,  $\widehat{\Delta}$ converges in probability to ${\Delta}$.
  \end{theorem}
\begin{theorem}
  As $n\rightarrow \infty$,  $\sqrt{n}(\widehat{\Delta}-\Delta)$ converges in distribution to normal random variable with mean zero and variance $\sigma^2$, where $\sigma^2$ is given by
  \begin{small}
 \begin{equation}\label{var}
\sigma^{2}=Var\Big(\frac{2X\bar F(X)}{\lambda} +\frac{2}{\lambda}\int_{0}^{X}ydF(y)+\small(1-k\small)X\int_{X}^{\infty}\frac{1}{y}
dF(y)+\frac{(1-k)}{X}\int_{0}^{X}ydF(y)\Big).
\end{equation}
\end{small}
\end{theorem}

\noindent {\bf Proof:}
Define $$\widehat{\Delta}^{*} =\frac{U_{1}}{\lambda}+(1-k)U_{2}.$$  Since $\widehat \lambda$ and $\widehat{k}$ are consistent estimators of $\lambda$ and $k$, respectively, by Slutsky's theorem,  the  asymptotic distribution of $\sqrt{n}(\widehat{\Delta}-\Delta)$  and  $\sqrt{n}(\widehat{\Delta}^{*}-E(\widehat{\Delta}^{*})$ are same. Now we observe  that $\widehat{\Delta}^*$ is a U statistic with symmetric kernel,
\begin{small}
$$h(X_1,X_2)  = \frac{1}{2}\left[\left(\frac{2\min(X_{1},X_{2})}{\lambda}\right)+
(1-k)\left(\frac{X_{1}}{X_{2}}I(X_{1}<X_{2})+\frac{X_{2}}{X_{1}}I(X_{2}<X_{1})\right)\right].$$
\end{small}
Hence using the central limit theorem for U-statistics  we have the asymptotic  normality of $ \widehat{\Delta}^*$. The asymptotic variance is $4\sigma_1^2$ where $\sigma_1^2$ is given by (Lee, 2019)
\begin{equation}\label{var1}
\sigma_1^2= Var\left[E\left(h(X_{1},X_{2})|X_{1}\right)\right].
\end{equation}Consider
\begin{eqnarray}\label{var11}
E[2\min(x,X_{2})]&=&2E[x I(x<X_{2})+X_2I(X_{2}<x)]\nonumber \\
&=&2xP(x<X_2)+2\int_{0}^{\infty}yI(y<x)dF(y)\nonumber\\
&=&2x\bar F(x)+2\int_{0}^{x}ydF(y).
\end{eqnarray}Also
\begin{eqnarray}\label{var12}
E[\frac{x}{X_2}I(x<X_2]+E[\frac{X_2}{x}I(X_2<x)] &=&xE[\frac{1}{X_2}I(x<X_2]+\frac{1}{x}E[X_2I(X_2<x)]\nonumber\\
&=&x\int_{0}^{\infty}\frac{1}{y}I(x<y)dF(y)+\frac{1}{x}\int_{0}^{x}ydF(y)\nonumber\\
&=&x\int_{x}^{\infty}\frac{1}{y}dF(y)+\frac{1}{x}\int_{0}^{x}ydF(y).
\end{eqnarray}
Substituting equations (\ref{var11}) and (\ref{var12})  in equation  (\ref{var1}) we obtain the variance expression as specified in the theorem.

Under the null hypothesis $H_0$, $\Delta{(F)}=0$. Hence we have the following corollary.
\begin{corollary}
 Under $H_0$, as $n\rightarrow \infty$,  $\sqrt{n}\widehat{\Delta}$ converges in distribution to normal with mean zero and variance $\sigma_0^2$, where $\sigma_0^2$ is the value of $\sigma^2$ evaluated under $H_0$.
  %\begin{small}
% \begin{equation}\label{var}
%\sigma^{2}=Var\Big(\frac{2X\bar F(X)}{\lambda} +\frac{2}{\lambda}\int_{0}^{X}ydF(y)+\small(1-k\small)X\int_{X}^{\infty}\frac{1}{y}
%dF(y)+\frac{(1-k)}{X}\int_{0}^{X}ydF(y)\Big).
%\end{equation}
%\end{small}
\end{corollary}

An asymptotic critical region of the test can be obtain  using Corollary 1. Let $\widehat\sigma_{0}^2$ be a consistent estimator of $\sigma_{0}^2$. We reject the null hypothesis $H_{0}$ against the alternative hypothesis $H_{1}$ at a significance level $\alpha$, if
\begin{equation*}
 \frac{ \sqrt{n} |\widehat{\Delta}| }{\widehat\sigma_0}>Z_{\alpha/2},
  \end{equation*}
where $Z_{\alpha}$ is the upper $\alpha$-percentile point of the standard normal distribution.  Since  the distribution function  $F$ has no closed form for the gamma distribution,  it is difficult to evaluate the null variance  $\widehat\sigma_{0}^2$ be a consistent estimator of $\sigma_{0}^2$. Hence we find the  critical region of the test based on Monte Carlo simulation. We determine lower ($c_1$) and upper ($c_2$)  quantiles in such a way that $P(\widehat\Delta<c_1)=P(\widehat\Delta>c_2)=\alpha/2$.  Finite sample performance of the  test is evaluated through Monte Carlo simulation study and the results are reported in Section 4.

\section{Test for right censored case}
Next we discuss how the censored observations can be incorporated in  the proposed testing procedure.
Consider the right-censored  data  $(Y, \delta)$, with $Y=\min(X,C)$ and $\delta=I(X\leq C)$, where $C$ is the censoring time. We assume censoring times and lifetimes are independent. Now we develop test  discussed  in Section 2 based on $n$ independent and identical observation $\{(Y_{i},\delta_i),1\leq i\leq n\}$.   Since we consider  U-statistics theory for right censored data, we use the same  departure measure $\Delta (F)$ given in (\ref{deltam}).  We rewrite (\ref{deltam}) as
\begin{eqnarray}\label{deltamcens}
\Delta(F)&=&\frac{1}{\lambda}E(\min(X_{1},X_{2}))+(1-k)E\left(\frac{X_{1}}{X_{2}}I(X_{1}<X_{2})\right)-kP(X_1<X_2).
\end{eqnarray}

To develop the test statistic for right censored case, we estimate each quantity in (\ref{deltamcens}) using U-statistics for right censored data (Datta et al., 2010). An estimator of $E(\min(X_{1},X_{2}))$ is given by
\begin{equation}\label{delta1c}
\widehat{\Delta}_{1c}=\frac{2}{n(n-1)}\sum_{i=1}^{n}\sum_{j<i;j=1}^{n}\frac{\min(Y_{1},Y_{2})\delta_i\delta_j}{\widehat{K}_{c}(Y_i)\widehat{K}_{c}(Y_j)},
\end{equation}
 provided $\widehat{K}_{c}(Y_i)>0$ and $\widehat{K}_{c}(Y_j)>0$, with probability 1 and $\widehat{K}_c$ is the Kaplan-Meier estimator of $K_c$, the survival function of  $C$.
 Again, an estimator of   $E\left(\frac{X_{1}}{X_{2}}I(X_{1}<X_{2})\right)$ is given by
 \begin{equation}\label{delta2c}
\widehat{\Delta}_{2c}=\frac{1}{n(n-1)}\sum_{i=1}^{n}\sum_{j<i;j=1}^{n}\frac{(\frac{X_{i}}{X_{j}}I(X_{i}<X_{j})+\frac{X_{j}}{X_{i}}I(X_{j}<X_{i}))\delta_i\delta_j}{\widehat{K}_{c}(Y_i)\widehat{K}_{c}(Y_j)}.
\end{equation}
An estimator of   $P(X_{1}<X_{2})$ is given by
 \begin{equation}\label{delta3c}
\widehat{\Delta}_{3c}=\frac{1}{n(n-1)}\sum_{i=1}^{n}\sum_{j<i;j=1}^{n}\frac{(I(X_{i}<X_{j})+I(X_{j}<X_{i}))\delta_i\delta_j}{\widehat{K}_{c}(Y_i)\widehat{K}_{c}(Y_j)}.
\end{equation}
Similarly, the estimators of $E(X)$ and $E(X^2)$ are given by (Datta, 2005)
 \begin{equation*}
\bar{X}_{c}=\frac{1}{n}\sum_{i=1}^{n}\frac{Y_i \delta_i}{\widehat{K}_{c}(Y_i)}.
\end{equation*}and
\begin{equation*}
\bar{X}_{c}^2=\frac{1}{n}\sum_{i=1}^{n}\frac{Y_i^2 \delta_i}{\widehat{K}_{c}(Y_i)}.
\end{equation*}Using  $\bar{X}_{c}$ and $\bar{X}_{c}^2$ we obtain the estimators of $k$ and $\lambda$. Moment estimators of $k$ and $\lambda$  are given by
$$\widehat k_m=\frac{E^2(X)}{Var(X)}$$ and
$$\widehat\lambda_m=\frac{Var(X)}{E(X)}.$$
 Hence under right censored case we estimate these quantities by
\begin{equation}\label{kc}
  \widehat{k}_c=\frac{(\bar{X}_{c})^2}{\bar{X}_{c}^2-(\bar{X}_{c})^2}.
\end{equation}and
\begin{equation}\label{lc}
  \widehat{\lambda}_c=\frac{\bar{X}_{c}^2-(\bar{X}_{c})^2}{\bar{X}_{c}}.
\end{equation}
Since $\bar{X}_{c}$ and $\bar{X}_{c}^2$ are consistent estimators of $E(X)$ and $E(X^2)$, we can easily verify that   $\widehat{k}_c$ and $\widehat{\lambda}_c$ are consistent estimators of  $k$ and $\lambda$, respectively.
Using the estimators given in equations (\ref{delta1c}-\ref{lc}), we obtain the test statistic in right censored case as
\begin{equation}\label{esticen}
  \widehat{\Delta}_{c}=\frac{\widehat{\Delta}_{1c}}{\widehat{\lambda}_{c}}+(1-\widehat{k}_c)\widehat{\Delta}_{2c}-{\widehat{k}_{c}}{\widehat{\Delta}_{3c}}.
\end{equation} We reject $H_{0}$ in favour of $H_{1}$ for large values of $\widehat{\Delta}_c$.

Next we obtain the limiting distribution of $ \widehat{\Delta}_{c}$. Let $N_i^c(t)=I(Y_i\leq t, \delta_i=0)$ be the counting process corresponds to the censoring variable $C_i$, $R_i(t)=I(Y_i\geq t)$. Also let $\lambda_c$ be the  hazard rate of $C$.  The martingale associated with this counting process $N_i^c(t)$ is given by
\begin{equation*}
M_i^c(t)=N_i^c(t)-\int_{0}^{t} R_i(u) \lambda_c(u) du.
\end{equation*}
Let $G(x,y)=P(X_{1}\leq x, Y_{1}\leq y,  \delta=1), x\in \mathcal{X}$, $\bar H(t)=P(Y_{1}> t)$ and
\begin{equation*}
w(t)=\frac{1}{\bar{H}(t)} \int_{\mathcal{X}\times[0,\infty)}{\frac{h_1(x)}{K_c(y-)}I(y>t)dG(x,y)},
\end{equation*}
where $h_1(x)=E(h(X_1,X_2|X_1=x)).$  The proof of next result follows from Theorem 1 of Datta et al. (2010) for a particular choice of the kernel.
\begin{theorem}\label{thm5.4}
Let \begin{small}
$$h_1(x)  =\frac{1}{2}E\left[\left(\frac{2\min(x,Y_{2})}{\lambda}\right)+
(1-k)\left(\frac{x}{Y_{2}}I(x<Y_{2})+\frac{Y_{2}}{x}I(Y_{2}<x)\right)-k\left(I(x<Y_{2})+I(Y_{2}<x)\right)\right].$$\end{small}
Suppose the conditions \begin{small} $$E\left[\frac{1}{2}\left(\left(\frac{2\min(Y_{1},Y_{2})}{\lambda}\right)+
(1-k)E\left(\frac{Y_{1}}{Y_{2}}I(Y_{1}<Y_{2})+\frac{Y_{2}}{Y_{1}}I(Y_{2}<Y_{1})\right)-k\right)\right]^2 <\infty,$$\end{small} $\int_{\mathcal{X}\times[0,\infty)}{\frac{h_1^{2}(x)}{K_c^2(y)}dG(x,y)}<\infty$ and  $\int_0^\infty w^2(t)\lambda_c(t)dt<\infty$ holds. As $n \rightarrow \infty $,  $\sqrt{n}(\widehat{\Delta}_c-\Delta)$ converges in  distribution to Gaussian random variable with mean zero and variance $4\sigma_{c}^{2}$, where $\sigma_{c}^2$  is given by
\begin{equation*}
\sigma_{c}^{2}=Var\Big(\frac{h_1(X)\delta_1}{K_c(Y_1-)}+\int w(t) dM_1^c(t)\Big).
\end{equation*}
\end{theorem}

Next we find an estimator of  $\sigma_{c}^{2}$ using the reweighed techniques. An estimator of $\sigma_{c}^2$ is given by
\begin{equation*}\label{ecvar}
\widehat{\sigma}_{c}^2=\frac{4}{(n-1)}\sum_{i=1}^{n}(V_{i}-\bar V)^2,
\end{equation*}
where
\begin{equation*}\label{36}
V_{i}=\frac{\widehat{h}_1(X_{i})\delta_i}{\widehat{K}_c(Y_{i})}+\widehat w(X_{i})(1-\delta_i)-\sum_{j=1}^{n}\frac{\widehat w(X_{i})I(X_{i}>X_{j})(1-\delta_i)}{\sum_{i=1}^{n}I(X_{i}>X_{j})},
\end{equation*}
$$\bar V =\frac{1}{n}\sum_{i=1}^{n}V_{i}, \quad  \widehat{h}_1(X)=\frac{1}{n}\sum_{i=1}^{n}\frac{h(X,Y_{i})\delta_i}{\widehat{K}_c(Y_{i}-)}, \quad R(t)=\frac{1}{n}\sum_{i=1}^{n}I(Y_{i}>t)$$ and
$$\widehat w (t)=\frac{1}{R(t)}\sum\limits_{i=1}^{n}\frac{\widehat{h}_1(X_{i})
\delta_{i}}{\widehat{K}_c(Y_{i})}I(X_{i}>t).$$

Let $\widehat{\sigma}_{0c}^2$ be the value of $\widehat{\sigma}_{c}^2$ evaluated under $H_0$. Under right censored situation, we reject the null hypothesis $H_{0}$ against the alternative hypothesis $H_{1}$ at a significance level $\alpha$, if
\begin{equation*}
 \frac{ \sqrt{n} |\widehat{\Delta}_c| }{\widehat{\sigma}_{0c}}>Z_{\alpha/2}.
  \end{equation*}
The results of the Monte Carlo simulation which assess the finite sample performance of the test  is also reported in Section 4.
\section{Empirical evidence}
To evaluate the finite sample performance of the proposed test procedure, we conduct a Monte Carlo simulation study using R software. To show the competitiveness of our test with the existing test procedures for complete data, we compare the empirical powers of the same. In censored case we evaluate the power of our test against different alternatives.

\subsection{Uncensored case}
We find the type I error and empirical power of the  proposed test and other acknowledged tests.  The algorithm used to find the empirical power can be summarised as follows.
\begin{enumerate}%\vspace{-0.1in}
\item Generate lifetime data from the desired alternative and calculate the test statistic.
\item Estimate the parameters $k$ and $\lambda$ from the data obtained in Step 1.
	\item Generate a sample of size $n$ from $\Gamma(k,\lambda)$.
	\item Obtain the bootstrap distribution of the test statistic  with 10000 bootstrap samples obtained from the data generated in Step 3 and determine the critical point.
	%\item Determine critical points from the bootstrap distribution.
	\item Repeat Steps 1-4 10000 times and calculate empirical power as the proportion of significant test statistics.
\end{enumerate}
First we find empirical type I error of the test.  We generate lifetimes from gamma distribution with different samples sizes $n=25,50,75, 100, 200$ to calculate the empirical type I error. To find the empirical  power, lifetime random variables are generated from different choices of alternative  including Weibull, lognormal and Pareto distributions where  the distribution functions are;\\
Weibull distribution: $F(x)=1-e^{(-x/\lambda)^{k}}$, $x>0$, $k,\,\lambda>0,$\\
Pareto distributions: $F(x)=(\lambda/x)^{\alpha}$ $x>0$, $\alpha,\, \lambda>0$,\\
Lognormal distribution: $F(x)=\Phi(\frac{\ln x-\mu}{\sigma})$, $x>0$, $-\infty<\mu<\infty,\, \sigma^2>0$,\\
where $\Phi(x)$ is the cumulative distribution function of the  standard normal random variable.
We compare the performance of our  test with  goodness of fit test for gamma distribution proposed by Henze et al. (2012)  and Betsch and Ebner (2019) and also with the well-known  Kolmogorov Smirnov (KS) test and Cramer von Mises (CvM) test.  The test statistic by Henze et al. (2012) is given by
\begin{equation*}
    HME=\int_{0}^{\infty}Z_{1n}^{2}(t)w(t)dt,
\end{equation*}
where $Z_{1n}(t)=\sqrt{n}\left[(1+t)L_n^{'}(t)+\widehat kL_n(t)\right]$ with $L_n(t)$ as the empirical Laplace transform defined as $L_n(t)=\frac{1}{n}\sum_{i=1}^{n}\text{exp}(-tY_j)$, $w(t)$ is a weight function satisfying some predefined conditions and $Y_i=\frac{X_i}{\widehat\lambda}, i=1,2,..,n$. Test statistic proposed by Betsch and Ebner (2019) based on fixed point characterization is given by
\begin{equation*}
    BE=\int_{0}^{\infty}Z_{2n}^{2}(t)w(t)dt,
\end{equation*} where
$Z_{2n}(t)=\sqrt{n} \left[\frac{1}{n}\sum_{i=1}^{n}(\frac{1-\hat k}{Y_i}+1)\text{min}(Y_i,t)-\frac{1}{n}\sum_{i=1}^{n}I(Y_i \le t)\right]$. The value of the Henze test  and Betsch and Ebner test are obtained using the R-Package `gofgamma'.
The Kolmogorov-Smirnov test statistic is given by $ KS=\text{max}\{D^{+}, D^{-}\}$ where
\begin{equation*}
  D^{+}=\max_{i=1,2,...,n}\left(\frac{i}{n}-\widehat F(X_{(i)})\right)~~~~\text {and}~~~~
  D^{-}=\max_{i=1,2,...,n}\left (\widehat F(X_{(i)})-\frac{i-1}{n}\right)
\end{equation*}
and Cramer von Mises statistics is given by
\begin{equation*}
    CM=\frac{1}{12n}+\sum_{i=1}^{n}\left (\widehat F(X_{(i)})-\frac{2i-1}{2n}\right)^{2},
\end{equation*}
where $\widehat F(.)$ is the empirical distribution function.
The results of the simulation study are given in Tables 1-4.  In Table 1, we report the empirical type I error of the proposed test and in all other tables, we give the empirical power  against different alternatives. From Table 1, we observe  that the size of the test attains chosen level of significance. The test has good power against all choices of alternatives which increases with sample size. We can see that the newly proposed test performs better than other tests in most of the cases we considered. When the alternative is  Pareto distribution, we note other test perform better than our test for small sample sizes.  The proposed test has  high power  even  for  small sample size, which affirms the efficiency of the test.

\begin{table}[h]
\caption{Empirical type I error}
\scalebox{0.8}{
\begin{tabular}{cccccccccccccc}\hline
\multirow{2}{*}{} & \multicolumn{2}{c}{$\widehat\Delta$} &\multicolumn{2}{c}{HME}&\multicolumn{2}{c}{BE}&\multicolumn{2}{c}{KS }&\multicolumn{2}{c}{CvM}  \\ \cline{1-11}
$n$ &   $\alpha=0.01$  & $\alpha=0.05$&   $\alpha=0.01$  & $\alpha=0.05$ & $\alpha=0.01$  & $\alpha=0.05$&   $\alpha=0.01$  & $\alpha=0.05$&$\alpha=0.01$  & $\alpha=0.05$ \\ \hline
25&  0.0104 &0.0494 &0.0013&0.0595&0.0130&0.0603&0.0095&0.0472 &0.0104 & 0.0470
\\
50& 0.0089&0.0491&0.0113& 0.0542&0.0116 & 0.0552&0.0115& 0.0510 &0.0101&  0.0530\\
75 &0.0089&0.0484 &0.0119 &0.0546&0.0117&0.0533 &  0.0092&0.0507&0.0099&0.0495\\
100 & 0.0097&0.0487&0.0114&0.0520& 0.0113& 0.0528&0.0102&0.0530&0.0097& 0.0534\\
200 & 0.0095&  0.0498& 0.0115&0.0541&0.0113&0.0529&0.0118 &0.0512& 0.0112& 0.0507\\ \hline
\end{tabular}}
\end{table}

\begin{table}[h]
\caption{Empirical power: Log normal distribution ($\mu=2,\,\sigma=1$) }
\scalebox{0.8}{
\begin{tabular}{ccccccccccccccc}\hline
\multirow{2}{*}{} & \multicolumn{2}{c}{$\widehat\Delta$}  &\multicolumn{2}{c}{HME}&\multicolumn{2}{c}{BE}&\multicolumn{2}{c}{KS }&\multicolumn{2}{c}{CvM}  \\\hline %\cline{1-13}
$n$ &   $\alpha=0.01$  & $\alpha=0.05$&   $\alpha=0.01$  & $\alpha=0.05$ & $\alpha=0.01$  & $\alpha=0.05$&   $\alpha=0.01$  & $\alpha=0.05$&$\alpha=0.01$  & $\alpha=0.05$ \\ \hline
25&1.0000&1.0000&0.1442 &0.3335& 0.1583&0.3322&1.0000&1.0000  &1.0000&1.0000   \\
50&1.0000&1.0000&0.4079& 0.6463 &0.4019&0.6218&1.0000&1.0000  &1.0000&1.0000 \\
75&1.0000&1.0000& 0.5738& 0.7991&0.5697 &0.7620&1.0000&1.0000 &1.0000&1.0000 \\
100&1.0000&1.0000&0.9951&1.0000&0.9930&0.9989&1.0000&1.0000 &1.0000&1.0000  \\
200&1.0000&1.0000&1.0000&1.0000& 1.0000& 1.0000& 1.0000&1.0000 & 1.0000&1.0000 \\\hline
\end{tabular}}
\end{table}
%
%\begin{table}[h]
%\caption{Empirical power the tests: Distribution 2}
%\scalebox{0.8}{
%\begin{tabular}{cccccccccccccc}\hline
%\multirow{2}{*}{} & \multicolumn{2}{c}{$\widehat\Delta$} &\multicolumn{2}{c}{Henze test-I}&\multicolumn{2}{c}{Henze test-II}&\multicolumn{2}{c}{KS test}&\multicolumn{2}{c}{AD test}&\multicolumn{2}{c}{CvM test}  \\ \cline{1-5}
%$n$ &   $\alpha=0.01$  & $\alpha=0.05$&   $\alpha=0.01$  & $\alpha=0.05$ & $\alpha=0.01$  & $\alpha=0.05$&   $\alpha=0.01$  & $\alpha=0.05$&$\alpha=0.01$  & $\alpha=0.05$ \\ \hline
%25& &&&&&&&&&\\
%50& &&&&&&&&&\\
%75&  &&&&&&&&&\\
%100& &&&&&&&&&\\
%200&  &&&&&&&&&\\ \hline
%\end{tabular}}
%\end{table}

\begin{table}[h]
\caption{Empirical power: Pareto distribution ($\alpha=2,\,\lambda=1$) }
\scalebox{0.8}{
\begin{tabular}{ccccccccccccccc}\hline
\multirow{2}{*}{} & \multicolumn{2}{c}{$\widehat\Delta$} &\multicolumn{2}{c}{HME}&\multicolumn{2}{c}{BE}&\multicolumn{2}{c}{KS }&\multicolumn{2}{c}{CvM}  \\\hline %\cline{1-13}
$n$ &   $\alpha=0.01$  & $\alpha=0.05$&   $\alpha=0.01$  & $\alpha=0.05$ & $\alpha=0.01$  & $\alpha=0.05$&   $\alpha=0.01$  & $\alpha=0.05$&$\alpha=0.01$  & $\alpha=0.05$ \\ \hline
25&0.8105& 0.8493&0.7132&0.9166&0.5918&0.8463& 1.0000& 1.0000 &1.0000& 1.0000  \\
50&0.8587& 0.9052 &0.9811&0.9981&0.9607&0.9974& 1.0000& 1.0000 &1.0000& 1.0000 \\
75& 0.8802& 0.9894&1.0000& 1.0000&1.0000& 1.0000& 1.0000& 1.0000 &1.0000& 1.0000\\
100&  0.9871& 1.0000&1.0000& 1.0000&1.0000& 1.0000& 1.0000& 1.0000 &1.0000& 1.0000\\
200& 1.0000& 1.0000&1.0000& 1.0000&1.0000& 1.0000& 1.0000& 1.0000 &1.0000& 1.0000\\\hline
\end{tabular}}
\end{table}

\begin{table}[h]
\caption{Empirical power: Weibull distribution ($k=2,\,\lambda=1$)}
\scalebox{0.8}{
\begin{tabular}{cccccccccccc}\hline
\multirow{2}{*}{} & \multicolumn{2}{c}{$\widehat\Delta$} &\multicolumn{2}{c}{HME}&\multicolumn{2}{c}{BE}&\multicolumn{2}{c}{KS }&\multicolumn{2}{c}{CvM}  \\\hline %\cline{1-13}
$n$ &   $\alpha=0.01$  & $\alpha=0.05$&   $\alpha=0.01$  & $\alpha=0.05$ & $\alpha=0.01$  & $\alpha=0.05$&   $\alpha=0.01$  & $\alpha=0.05$&$\alpha=0.01$  & $\alpha=0.05$ \\ \hline
25&0.8424&	0.9856&	0.0426&	0.1344&	0.1130&	0.1139&	0.1147&	0.3836&	0.0780&	0.4192\\
50&1.0000&	1.0000&	0.0862&	0.2352&	0.1972&	0.1972&	0.4962&	0.8252&	0.5622&	0.9236\\
75&1.0000&	1.0000&	0.1391&	0.3234&	0.2625&	0.2621&	0.8188&	0.9768&	0.9028&	0.9967\\
100&1.0000&	1.0000&	0.1836&	0.3772&	0.3188&	0.3188&	0.9539&	0.9980&	0.9873&	1.0000\\
200&1.0000&	1.0000&	0.4204&	0.6608&	0.5742&	0.5744&	1.0000&	1.0000&	1.0000&	1.0000\\\hline
\end{tabular}}
\end{table}

\subsection{Censored case}

We calculate empirical type I error and power of the test statistic proposed for right censored data using Monte Carlo simulation studies. To calculate the empirical type I error, lifetimes are generated from gamma distribution. We considered the same alternatives as in uncensored case for finding  the empirical power. Here, the censoring  percentages  are chosen to be 20\% and 40\%.  In all cases, the censoring random variable $C$ is generated from exponential distribution with parameter $b$,  where $b$ is chosen such that $P(T>C)=0.2(0.4)$.  Re-weighting techniques explained in Section 3 is used to estimate the variance of $\widehat\Delta_c$. Results of the simulation study are presented in Tables 5 and 6.

%\begin{enumerate}
	
%	\item Generate random samples from the desired distribution and estimate %	\item To calculate the critical values of the test, first generate censoring random variables $C_1^{'},C_2^{'},...,C_n^{'}$ from the survival distribution $\hat K_c(.)$.
%	\item Generate new samples $Y_1^{'},Y_2{'},..., Y_n{'}$ from gamma distribution specified in $H_0$.
%	\item Using the random variables generated in Step 2 and 3, construct a bootstrap sample  of $(Y_i^{b},\delta_i^{b})$ where $Y_i^{b}=min(Y_i^{'},C_i^{'})$ and $\delta_i^{b}=I(Y_i^{'} \le C_i^{'})$ for $i=1,2,..,n$.
%	\item Based on $(Y_i^{b},\delta_i^{b})$, calculate the test statistic and repeat the procedure $B$ times.
%	\item   Obtain the boot strapped distribution of the test static and estimate critical values for chosen significance level.
%	\item Reject $H_0$, if the test statistic falls in critical region.
%	\item Repeat the process 10000 times and estimate power of the test as  percentage of rejected $H_0$.
%\end{enumerate}

%...................20% censoring
\begin{table}[h]
\caption{Empirical type I error and power of the test when 20\% of lifetimes are censored. }
\scalebox{0.8}{
\begin{tabular}{cccccccccccccc}\hline
\multirow{2}{*}{} & \multicolumn{2}{c}{Gamma (1,1)} & \multicolumn{2}{c}{Lognormal (2,1)}&\multicolumn{2}{c}{Weibull (2,1)}&\multicolumn{2}{c}{Pareto (2,1)} \\ \hline
$n$ &   $\alpha=0.01$  & $\alpha=0.05$&   $\alpha=0.01$  & $\alpha=0.05$ & $\alpha=0.01$  & $\alpha=0.05$&   $\alpha=0.01$  & $\alpha=0.05$ \\ \hline
50 &0.0132&0.0461 &0.9863&0.9916&0.9891&0.9948&0.8381&0.8892\\
75& 0.0121&0.0489 &0.9992&1.0000&1.0000&1.0000&0.8675&0.9543\\
100& 0.0118&0.0508&1.0000&1.0000&1.0000&1.0000&0.9782&0.9999\\
200&  0.0104&0.0497&1.0000&1.0000&1.0000&1.0000&1.0000&1.0000\\ \hline
\end{tabular}}
\end{table}
%....40% censoring
\begin{table}[h]
\caption{Empirical type I error and power of the test when 40\% of lifetimes are censored. }
\scalebox{0.8}{
\begin{tabular}{cccccccccccccc}\hline
\multirow{2}{*}{} & \multicolumn{2}{c}{Gamma(1,1)} & \multicolumn{2}{c}{Lognormal (2,1) }&\multicolumn{2}{c}{Weibull (2,1)}&\multicolumn{2}{c}{Pareto (2,1)}\\\hline
$n$ &   $\alpha=0.01$  & $\alpha=0.05$&   $\alpha=0.01$  & $\alpha=0.05$ & $\alpha=0.01$  & $\alpha=0.05$&   $\alpha=0.01$  & $\alpha=0.05$\\ \hline
50 &0.0140&0.0514&0.5248&0.6069&0.9594&0.9847& 0.7893 &0.8456\\
75 &0.0112& 0.0489&0.6785&0.7494&0.9976&0.9995&0.8175&0.8893\\
100&0.0108&0.0507 &0.8044&0.8798&1.0000&1.0000&0.9274&0.9632\\
200&0.0106&0.0505 &0.9627&0.9778&1.0000&1.0000&0.9461&0.9874\\ \hline
\end{tabular}}
\end{table}

We can see that the  empirical power of the test approaches  the chosen level significance  as $n$ increases. The performance of the test is good in terms of empirical power. From Tables 5 and 6, we observe that the power of the test  increases with sample size and decreases with censoring percentage.
\section{Data analysis}
The proposed test procedures are illustrated using several real data sets.
\subsection{Complete data}
We consider two data sets for the analysis. To find the critical region, we use  the following  algorithm.
\begin{enumerate}%\vspace{-0.1in}
	\item Estimate the parameters of gamma distribution $k$ and $\lambda$ from the observed data.
	\item Generate a random sample from gamma distribution  using the estimated parameters in Step 1.
	\item Obtain the bootstrap distribution of the test statistic with 10000 bootstrap samples from the data generated in Step 2 and determine the critical points.
\end{enumerate}
Illustration 1: We consider the data on survival times in weeks for 20 male rats
that were exposed to a high level of radiation. The data is discussed in Lawless (2011) in Example 4.2.1. We choose a data with small sample size to make sure that our test is suitable for any sample size. The test statistic is obtained as $0.0964$  where the $5\%$ level critical values are  0.0456 and  1.2664
 and $1\%$ level critical values are  -0.0055 and   2.1042
respectively.  Hence we accept $H_0$ that the data on  survival times of male rats follows gamma distribution.

\noindent Illustration 2 : We also examine the lifetime data on number of millions of revolutions before failure for 23 ball bearings. The data is studied in Lawless (2011) in context of goodness of fit tests for parametric models and given in Example 3.3.1. We obtain the test statistic as -0.0379 where the $5\%$ level critical values are  0.0956 and 1.2978 and $1\%$ level critical value are 0.0385 and  2.0577. Hence we reject the null hypothesis that the data follows gamma distribution.
\subsection{Censored data}
Two real data sets are considered for illustrating  the  proposed test procedure. We use the normal based critical region given in Section 3 to make a decision.   The asymptotic null variance of $\widehat\Delta_c$ is estimated using the re-weighting techniques explained in Section 3. \\
Illustration 1: We analyse stanford heart transplant data available in R software  named `stanford2' to test for gamma assumption. The data consist of 184 lifetimes where 72 of them are censored lifetimes. The censoring percentage of data is 38.5\%. The test statistic is calculated as 0.3503. So we accept the null hypothesis that the data follows gamma distribution at both 1\% and 5\% level of significance.

\noindent Illustration 2: We examine the data on lifetimes of disk break pads on 40 cars studied in Lawless (2011). Complete data set is given in Table 6.11, Page 337 of Lawless (2011). Out of the 40 observed lifetimes, 9 are censoring times, hence data contains 22.5\% of censored observations. The test statistic is obtained as 3.5142. Hence we reject the hypothesis of gamma distribution assumption for this data at both 1\% and 5\% level of significance.
\section{Concluding Remarks}
Based on fixed point characterization arising from Stein's type identity, we developed new goodness of fit test for gamma distribution. We studied the asymptotic properties of the proposed test statistic. The proposed test has well controlled error rate.  The power of the test  is compared with recently developed test for gamma distribution. The proposed test is illustrated using two real data sets.

Even though several tests are available for gamma distribution  in literature, as our knowledge,  all of these test are developed for complete data. Motivated by this we develop a new  goodness of fit test for gamma distribution with right censored data. We prove that the asymptotic distribution of the proposed test statistic is normal. We also find a consistent estimator of the asymptotic variance.  The finite sample performance of the test is evaluated through Monte Carlo simulation study.  Apart from right censoring, truncation and other types of censoring are common in lifetime data analysis. The proposed test can be modified to incorporates these situations. The similar goodness of fit test can be developed for other lifetime distribution using Stein's type characterization.

\section*{Acknowledgements}
Vaisakh  K. M. and Sreedevi E. P.  would like to thank Kerala State Council for Science, Technology and Environment for the financial support to carry out this research work.


\begin{thebibliography}{xx}
%\bibitem{} AbouRizk, S. M., Halpin,  D. W. and Wilson, J. R. (1994). Fitting beta distributions based on sample
%data. {\em Journal of Construction Engineering and Management}, 120, 288–-305.

%\bibitem{AP2017}Allison, J. S. and Pretorius, C. (2017). A Monte Carlo evaluation of the performance of two new tests for symmetry. {\em Computational Statistics,} 32, 1323--1338.

%\bibitem{AD} Anderson, T. W. and Darling, D. A. (1954). A test of goodness of fit. {\em Journal of the American Statistical Association,} 49, 765--769.

\bibitem{bhaumik2009testing}
Bhaumik, D. K., Kapur, K. and Gibbons, R. D. (2009). Testing parameters of a gamma distribution for small samples. {\em Technometrics}, 51, 326-334.

\bibitem{Barlow1996}
Barlow, R. E. and  Proschan, F. (1996). {\em Mathematical Theory of Reliability}. Society for Industrial and Applied Mathematics.

\bibitem{} Betsch, S. and Ebner, B. (2019). A new characterization of the gamma distribution and associated goodness-of-fit tests. {\em Metrika}, 82, 779--806.
\bibitem{}
Baringhaus, L., Ebner, B. and Henze, N. (2017). The limit distribution of weighted $L^2$ goodness-of-fit statistics under fixed alternatives, with applications. {\em Annals of the Institute of Statistical Mathematics}, 69, 969--995.
\bibitem{}
Datta, S., Bandyopadhyay, D. and  Satten, G. A. (2010). Inverse probability of censoring weighted u‐statistics for right‐censored data with an application to testing hypotheses. {\em Scandinavian Journal of Statistics}, 37, 680--700.

\bibitem{}
Deshpande, J. V. and  Purohit, S. G. (2015). {\em Lifetime Data: Statistical Models and Methods}. World Scientific Publishing Company, Singapore.
\bibitem{}
Henze, N., Meintanis, S. G. and Ebner, B. (2012). Goodness-of-fit tests for the gamma distribution based on the empirical Laplace transform. {\em Communications in Statistics-Theory and Methods}, 41, 1543--1556.

\bibitem{}
Husak, G. J., Michaelsen, J. and  Funk, C. (2007). Use of the gamma distribution to represent monthly rainfall in Africa for drought monitoring applications. {\em International Journal of Climatology: A Journal of the Royal Meteorological Society}, 27, 935--944.
\bibitem{}
Kallioras, A. G., Koutrouvelis, I. A. and  Canavos, G. C. (2006). Testing the fit of gamma distributions using the empirical moment generating function. {\em Communications in Statistics—Theory and Methods}, 35, 527--540.

\bibitem{}Lawless, J. F. (2011). {Statistical Models and Methods for Lifetime Data}. John Wiley and Sons, New Jersey.

\bibitem{} Lehmann, E. L. (1951). Consistency and unbiasedness of certain nonparametric tests.{\em  The Annals of Mathematical Statistics}, 22, 165--179.

\bibitem{} Lee, A. J. (2019).  \textit{U-Statistics: Theory and Practice}, Marcel Dekker Inc., New York.

\bibitem{} Sudheesh, K. K. (2009). On Stein's identity and its applications. {\em Statistics \& Probability Letters,} 79, 1444--1449.

\bibitem{}Sudheesh, K. K. and  Dewan, I.  (2016). On generalized moment identity and its applications: A unified approach. {\em Statistics,} 50, 1149--1160.

\bibitem{}
Stein, C. (1972). A bound for the error in the normal approximation to the distribution of a sum of dependent random variables. {\em In Proceedings of the sixth Berkeley symposium on mathematical statistics and probability}, 2, 583--602.

\bibitem{}
Villaseñor, J. A. and González-Estrada, E. (2015). A variance ratio test of fit for Gamma distributions. {\em Statistics \& Probability Letters}, 96, 281--286.




\end{thebibliography}
\end{document}